\documentclass[aps,pra,twocolumn,superscriptaddress,eqsecnum]{revtex4}
\usepackage{epsfig}
\usepackage{graphicx}
\usepackage{bm}
\begin{document}

\title
{Entanglement transformation with no classical communication}
\author{Bing He} 
\email{bhe98@earthlink.net}
\affiliation{Department of Physics and Astronomy, Hunter College of the City University of New York, 695 Park Avenue, New York, NY 10065, USA}
\author{J\'{a}nos A. Bergou}
\affiliation{Department of Physics and Astronomy, Hunter College of the City University of New York, 695 Park Avenue, New York, NY 10065, USA}

\date{\today}
\begin{abstract}
We present an optimal scheme to realize the transformations between single copies of two bipartite entangled states without classical communication between the sharing parties. The scheme achieves the upper bound for the success probabilities 
[\pra {\bf 63}, 022301 (2001), \prl {\bf 83}, 1455 (1999)] of generating maximally entangled states if applied to entanglement concentration. Such strategy also dispenses with the interaction with an ancilla system in the implementation. We also show that classical communications are indispensable in realizing the deterministic transformations of a single bipartite entangled state. With a finite number of identical pairs of two entangled bosons, on the other hand, we can realize the deterministic transformation to any target entangled state of equal or less Schmidt rank through an extension of the scheme.

\end{abstract}


\maketitle

\section{Introduction} \label{section1}

Entanglement is a key resource in realizing various quantum information protocols, as well as manifesting the non-classical nature of physics. An entangled quantum state produced in laboratory and transported over a distance inevitably shows some degradation (the fidelity less than one) due to decoherence and loss in the process. To faithfully teleport a quantum state to a remote place, it is necessary to have a pair of particles in the maximally entangled (ME) state \cite{teleport}. 
This problem gives rise to the study of entanglement concentration \cite{bennett}, which is the procedure to generate an 
ME state out of copies of partially entangled states. 

One of the generalized problems of this type is how to realize the transformation between two entangled pure states with 
the possible supply of multiple copies or even a single copy \cite{nielsen, vidal} of the input state. An arbitrary input state of this type shared by two parties, say Alice and Bob, is given as $|\Psi_1\rangle=\sum_{i,j}\lambda_{i,j}|i\rangle_A|j\rangle_B$ (normalized), where the set $\{|1\rangle_A,\cdots,|M\rangle_A\}$ forms an orthonormal basis of Alice's space ${\cal H}_A$ and $\{|1\rangle_B,\cdots,|N\rangle_B\}$ forms that of Bob's space ${\cal H}_B$ ($M$ and $N$ can be different). 
A routine strategy to transform a single copy of this state involves a generalized measurement of Alice followed by 
one-way communication of its result to Bob \cite{nielsen, vidal, lo-popescu}. Classical communications coordinate the operations of Alice and Bob. It has been concluded that strategies with at least one-way classical communication are more powerful than those with no communication in manipulating one bipartite entangled state \cite{lo-popescu}. Against intuitions, classical communication is not cheap in some situations \cite{lo-popescu2}. A typical example is the superdense coding \cite{dense-coding}: Alice and Bob can use $n$ qubits to transmit $2n$ bits of classical information, if they share 
$n$ pairs of Bell states. However, if the entangled pairs they share are not perfect and it costs them more than $n$ bits of classical communication to transform the pairs to the ME states, it would totally destroy the purpose of superdense coding. In the collective multi-copy concentration of a bipartite entangled state
(the Schmidt projection of Bennett, {\it et al.} \cite{bennett} and its development \cite{u-concentration}), a classical communication channel is not necessary. In the case of single copy, however, it was an open question to find the optimal
strategy without communication to concentrate a general bipartite pure state to an ME state \cite{lo-popescu}. 

In this work we present an approach to transforming a bipartite entangled state only through the local operations of Alice or Bob (without the need for a classical communication channel), which is also equivalent to only transforming the reduced density matrices $\rho_A=Tr_B|\Psi_1\rangle\langle\Psi_1|$ or $\rho_B=Tr_A|\Psi_1\rangle\langle\Psi_1|$. 
By introducing the {\it coefficient matrix} $\Lambda=(\lambda_{i,j})$ \cite{notation,matrix}, 
where $\sum_{i,j}|\lambda_{i,j}|^2=1$, we rewrite the input state as
\begin{eqnarray}
|\Psi_1\rangle=(|1\rangle_A,\cdots,|M\rangle_A)\left(\begin{array}{ccc} \lambda_{1,1}& \cdots &\lambda_{1,N} \\
\vdots &\ddots &\vdots \\
\lambda_{M,1} & \cdots & \lambda_{M,N}
\end{array}\right)\left(\begin{array}{c}|1\rangle_B\\
 \vdots\\
 |N\rangle_B\\
 \end{array}\right).~~~~
\label{a}
\end{eqnarray}
It is straightforward that any operation to transform $|\Psi_1\rangle=\alpha^T\Lambda\beta$ by the individual effort of 
Alice or Bob is equivalent to the local manipulation of their basis vectors $\alpha=(|1\rangle_A,\cdots,|M\rangle_A)^T$ or $\beta=(|1\rangle_B,\cdots,|N\rangle_B)^T$. One example of such local manipulation on their bases is that Alice and Bob independently apply the operations $\alpha^T\rightarrow {\alpha}^TU^{\dagger}$ and ${\beta}\rightarrow V{\beta}$ with the unitary operators obtained from the singular value decomposition $\Lambda=U^{\dagger}\Lambda_dV$ and transform the coefficient matrix of $|\Psi_1\rangle$ to a diagonal form
\begin{eqnarray}
\Lambda_d
=diag(\lambda_1, \lambda_2, \cdots, \lambda_{min(M,N)}),
\end{eqnarray}
where $min(M,N)$ is the Schmidt rank of $|\Psi_1\rangle$. This diagonal matrix is abbreviated as $\lambda_i\delta_{i,j}$. Then the essential step in transforming the state to another entangled state $|\Psi_2\rangle$ is reduced to the transformation between their different coefficient matrices.
Here $K^{\dagger}$, $K^T$, $K^{\ast}$ represent the Hermitian conjugate, the transpose and the entry-wise complex conjugate of an arbitrary matrix $K$, respectively. The Schmidt coefficients $\lambda_i$ can be arranged in a standard form with $\lambda_i\geq\lambda_{i+1}$ by permuting the basis vectors. 
With the coefficient matrix $\Lambda$, the reduced density matrices of a bipartite entangled state are given 
as $\rho_A=\Lambda\Lambda^{\dagger}$ and $\rho_B=\Lambda^{T}\Lambda^{\ast}$.
Two permissible linear maps $A$ and $B$ from Alice and Bob, respectively, can be expressed as $A\otimes B|\Psi\rangle\rightarrow (\alpha')^T A\Lambda B^T \beta'$. 
 
\section{Transformation strategy in extended space} \label{section2}

Without loss of generality, we might ask Alice to work alone on her part of $|\Psi_1\rangle$ toward the target state $|\Psi_2\rangle$. $|\Psi_1\rangle$ and $|\Psi_2\rangle$ are assumed with the diagonal coefficient matrices $\Lambda_d=\lambda_i\delta_{i,j}$ and $\Sigma_d=\sigma_i\delta_{i,j}$, respectively, since we have shown that they 
can be transformed to this form with the uncorrelated local unitary operations.
To relate these two coefficient matrices, Alice can choose an operator $A=\cos \theta_i\delta_{i,j}$ such that $\cos\theta_i\lambda_i=\sqrt{p}~\sigma_i$ or $A\Lambda_d=\sqrt{p}~\Sigma_d$ in the matrix form, where $0\leq\theta_i< \frac{\pi}{2}$, for $i=1,\cdots,min(M,N)$, are the independent parameters and $p<1$ the common factor. Such linear transformation with $||A||\leq1$ is termed {\it contraction} \cite{bt}.

To implement a local transformation that changes the Schmidt coefficients of an entangled state, such as the above non-unitary transformation $A$ which corresponds to an element of a positive operator valued measurement (POVM), we could apply an ancilla and perform a joint unitary transformation on the tensor product space of the system and the ancilla (see, e.g., \cite{lo-popescu}). Here we adopt a different approach that saves the interaction with an ancilla. The point is that $|\Psi_1\rangle$ expressed in the form of Eq. (\ref{a}) can be rewritten 
as
\begin{eqnarray}
|\Psi_1\rangle=\alpha^T \Lambda_d \beta=(\alpha^T, \alpha_c^T)\left(\begin{array}{cc} \Lambda_d \\
0 
\end{array}\right)\beta 
\end{eqnarray}
by extending Alice's space ${\cal H}_A$ to ${\cal H}_{A}\oplus {\cal H}_{A}$ with a complementary basis
vector $\alpha_c=(|M+1\rangle_A,\cdots,|2M\rangle_A)^T$. The $M\times N$ sub-matrix $(0)$ here represents that with all entries $0$. 
In this extended space we construct a unitary operator \cite{bt}
\begin{eqnarray}
U_{0}=\left(\begin{array}{cc} A &-(I-AA^{\dagger})^{1/2}  \\
(I-A^{\dagger}A)^{1/2} & A^{\dagger}
\end{array}\right),
\label{U}
\end{eqnarray}
which is a $2M \times 2M$ matrix. Under $U_{0}$ the basis vector $(\alpha^T, \alpha_c^T)$ is mapped to $(\alpha^T, \alpha_c^T)U^{\dagger}_{0}=(\alpha_{1'}^T, \alpha_{2'}^T)$, effecting the following transformation of the input $|\Psi_1\rangle$:
\begin{eqnarray}
|\Psi_{1}\rangle&\rightarrow& (\alpha_{1'}^T, \alpha_{2'}^T)~U_0\left(\begin{array}{cc} \Lambda_d \\
0 
\end{array}\right)\beta \nonumber\\
&=& \alpha_{1'}^TA\Lambda_d~ \beta+\alpha_{2'}^T(I-A^{\dagger}A)^{1/2}\Lambda_d~ \beta\nonumber\\
&=& \sqrt{p}~|\Psi_2\rangle+\sqrt{1-p}~|\Psi_f\rangle=|\Psi_{out}\rangle, 
\end{eqnarray}
where $|\Psi_f\rangle=(1-p)^{-1/2}\alpha_{2'}^T(I-A^{\dagger}A)^{1/2}\Lambda_d\beta$.
Alice can simultaneously perform two projections $P_{\alpha_{1'}}=\sum_{i'=1}^{M}|i'\rangle_A\langle i'|$ and 
$P_{\alpha_{2'}}=\sum_{i'=M+1}^{2M}|i'\rangle_A\langle i'|$, with $P_{\alpha_{1'}}+P_{\alpha_{2'}}=I_{2M}$, on the output 
$|\Psi_{out}\rangle$ because $[P_{\alpha_{1'}}, P_{\alpha_{2'}}]=0$.
If the operation $P_{\alpha_{1'}}\otimes I_B$ on $|\Psi_{out}\rangle$ succeeds, the target state $|\Psi_2\rangle$ will be 
projected out with a probability $p$. This unilateral operation strategy 
can be easily implemented for pairs of entangled photons, since the transformations on the photonic mode vector $\alpha^T=(a^{\dagger}_1,\cdots,a^{\dagger}_M)$ can be realized with linear optical circuits \cite{reck, bhe} and the projections on photonic modes can be done with quantum non-demolition (QND) measurement \cite{QND}. 

To maximize this success probability $p$, Alice should produce in the transformation the Schmidt coefficients $\sigma_i$ of the target state in an exact descending order, too \cite{notes1}. The maximum success probability is then given as 
\begin{eqnarray}
p^{opt}=min_k~max_{\{0\leq x_k\leq 1\}}~ x_k\lambda^2_k/\sigma^2_k,
\end{eqnarray}
where $x_k$, which are in the range of $\cos^2\theta_k$, can simultaneously take the value $1$ in the maximization since they are all independent. The optimum transformation probability of $U_{0}$ is therefore $p^{opt}=min_k \lambda^2_k/\sigma^2_k$. With this optimum success probability, Alice can always find the proper
parameters $\cos\theta_i=(\frac{\sigma_i}{\lambda_i})/(max_k\frac{\sigma_k}{\lambda_k})\leq1$ for the unitary transformation $U_{0}$. 

We recall some features of this transformation strategy: (1) The unilateral unitary transformations in the extended space are equivalent to a POVM with two elements $\{A^{\dagger}A, I-A^{\dagger}A\}$ performed in each ${\cal H}_A$ respectively:
\begin{eqnarray}
U_0\left(\begin{array}{cc} \rho_A & 0\\
0 & 0
\end{array}\right) U^{\dagger}_0=\left(\begin{array}{cc} A\rho_A A^{\dagger}& A\rho_A\tilde{A}\\
\tilde{A}\rho_AA^{\dagger} & \tilde{A}\rho_A \tilde{A}
\end{array}\right), 
\end{eqnarray}
where $\tilde{A}=(I-A^{\dagger}A)^{1/2}$. The strategy is optimized with $Tr(\rho_A A^{\dagger}A)$ being the maximum. After reaching this maximum, Alice cannot extract any $|\Psi_2\rangle$ from the residual part spanned by $\alpha_{2'}\otimes\beta$ because $min_i~\sin^{2}\theta_i~\lambda^2_i/\sigma^2_i=0$; (2) Under a bilateral action together with Bob performing a unitary operation in his extended space, which is constructed with another contraction $B$, an entangled state with the coefficient matrix $\Lambda_d$ will evolve to 
\begin{eqnarray}
|\Phi\rangle&=&\alpha^T_{1'}A\Lambda_dB^{T}\beta_{1'}+\alpha^T_{1'}A\Lambda_d(I-B^TB^{\ast})^{1/2}\beta_{2'}\nonumber\\
&+&\alpha^T_{2'}(I-A^{\dagger}A)^{1/2}\Lambda_dB^{T}\beta_{1'}\nonumber\\
&+&\alpha^T_{2'}(I-A^{\dagger}A)^{1/2}\Lambda_d (I-B^TB^{\ast})^{1/2}\beta_{2'},
\end{eqnarray}
which is the superposition of four orthogonal bipartite states. The transformation efficiency to a target will not be increased by making one of them proportional to $|\Psi_2\rangle$, since its amplitude is reduced by the other contraction 
$B$ and the remainder is distributed in other three components; 
(3) Applied for concentrating the entanglement of a pure state, this transformation strategy can produce an m-ME state $\frac{1}{\sqrt{m}}\sum_i|i\rangle_A|i\rangle_B$, where $m\leq min(M,N)$, with an optimum success probability $m\lambda^2_m$.
It saturates the entanglement concentration upper bound provided by Lo \& Popescu \cite{lo-popescu} if we concentrate the 
input $|\Psi_1\rangle$ to an ME state of the same Schmidt rank, and is also equal to or larger than Jonathan \& Plenio's distributed optimum probabilities \cite{J-P} to concentrate $|\Psi_1\rangle$ of rank $min(M,N)$ to an ensemble of 
ME states with equal and smaller Schmidt ranks. On the other hand, this scheme realizes the optimum concentration or 
dilution of a bipartite entangled pure state without a joint unitary transformation with an ancilla. 

\section{Necessity of classical communication in deterministic transformation} \label{section3}

The extensions of this unilateral strategy can be found in other interesting applications. One is the realization of the deterministic transformations of bipartite entangled states \cite{nielsen}. 
As we show in what follows, a classical communication channel is indispensable in such transformations. A deterministic transformation is realizable only if the Schmidt coefficient vector of the input $|\Psi_1\rangle$ is majorized by that of 
the target state $|\Psi_2\rangle$. Then, we will be able to find the doubly stochastic matrix to transform the vector with the entries of the squared Schmidt coefficients of $|\Psi_2\rangle$ to that of $|\Psi_1\rangle$, leading toward a local POVM which realizes $A_i\rho_{A_1}A^{\dagger}_i=p_i\rho_{A_2}$ (up to a unitary transformation) by each element $A_i$ \cite{j-s}. $\rho_{A_1}$ and $\rho_{A_2}$ here are, respectively, the reduced density matrices of $|\Psi_1\rangle$ and $|\Psi_2\rangle$ on Alice's side, and the probabilities satisfy $\sum_ip_i=1$.
The action of this POVM is equivalent to that of a unitary operator,
\begin{eqnarray}
U_{1}=\left(\begin{array}{ccc} A_1 & \cdots & \cdots  \\
\vdots & \vdots &\vdots \\
A_n &\cdots &\cdots 
\end{array}\right),
\end{eqnarray}
performed on the extended space spanned by $\alpha^T=(\alpha_{1}^T, \cdots, \alpha_{n}^T)$, where $\alpha_i=(|(i-1)M+1\rangle_A,\cdots, |iM\rangle_A)^T$.
The first $M$ columns of this matrix are orthonormal because $\sum_iA^{\dagger}_iA_i=I$, so they can be extended to a unitary matrix. The effect of $U_1$ is that the input state will be transformed as follows:
\begin{eqnarray}
|\Psi_1\rangle &\rightarrow & (\alpha_{1'}^T, \cdots, \alpha_{n'}^T)~U_1\left(\begin{array}{c} \Lambda_d \\
0 
\end{array}\right)\beta \nonumber\\
&=& (\alpha_{1'}^T, \cdots, \alpha_{n'}^T)\left(\begin{array}{c} \sqrt{p_1}~V_1\Sigma_d V_1\\
\vdots\\
\sqrt{p_n}~V_n\Sigma_d V_n 
\end{array}\right)\beta,
\end{eqnarray}
where $\Sigma_d$ is the coefficient matrix of the target state $|\Psi_2\rangle$, $(0)$ the submatrix of $(n-1)M\times N$ dimension, and $V_i$ the permutation matrices on the respective basis vectors of Alice and Bob \cite{j-s}. 
All $V_i$ here cannot be identical, otherwise the reduced density matrices $\rho_{A_1}=\Lambda_d\Lambda_d^{\dagger}$ and $\rho_{A_2}=\Sigma_d\Sigma_d^{\dagger}$ will have the same eigenvalue spectrum. 
With the projections $P_{i'}$ that can be implemented simultaneously by Alice, each $V_i\Sigma_d V_i$ can be projected out with a probability $p_i$. After she obtains the measurement result $1$ from one of $P_{i'}$,
Alice will need to communicate the result to Bob, indicating which $V_i$ he should perform to realize 
the target state. Based on the above procedure, we can also realize the optimal probabilistic transformation between
an arbitrary pair of entangled states with the equal Schmidt ranks in \cite{vidal}.

\section{Deterministic transformation with finite input copies}\label{section4}

The second extension is to realize the deterministic transformation to any target state, if provided with finite copies of 
a state $|\Psi_1\rangle$ with the same Schmidt rank as the target. These copies of entangled bosons \cite{boson} are prepared as the input
\begin{eqnarray}
|\Psi_{in}\rangle=\prod_{i=1}^n|\Psi_{1,i}\rangle=\prod_{i=1}^n\alpha^T_{i}\Lambda_d~\beta_{i},
\end{eqnarray}
where the basis vectors are given as $\alpha_{i}=(|1\rangle_{A_i},\cdots,|M\rangle_{A_i})^T$ and $\beta_{i}=(|1\rangle_{B_i},\cdots,|N\rangle_{B_i})^T$. From each copy Alice can produce through a previously discussed entanglement transformation an unnormalized state $\sqrt{p_i}|\Psi_{2,i}\rangle=\sqrt{p_i}\alpha_{i}^T\Sigma_d\beta_{i}$ 
in one subspace, if she dose not perform the final projection to the target. The probability distribution $\{p_i\}$ of these states can be tailored to satisfy $\sum^n_{i=1}p_i=1$. 
In each piece of such product, Alice's particle can be regarded as fractional 
\cite{particle}, so she could place the basis vectors $\alpha_i$ into the orthogonal subspaces and transform them as a direct sum. If $|\Psi_1\rangle$ is a separable state with a rank one $\Lambda_d$, this operation on the outputs of $n$ basic entanglement transformations is the inverse process of a single particle state being mapped to identical ones with some different probabilities $p_i$, which can be realized by the unitary operation
\begin{eqnarray}
U_{2}=\left(\begin{array}{ccc} \sqrt{p_1}~I & \cdots & \cdots  \\
\vdots & \vdots &\vdots \\
\sqrt{p_n}~I &\cdots &\cdots 
\end{array}\right).
\end{eqnarray}

The same process can be also implemented if the input is an entangled one, since Alice is performing the local operations only. In this case, the transitional state of the $(1+n)$-particle system after Alice has processed $n$ input copies is constructed with the correlation and the symmetry between the untouched particles of Bob. We define a set of vectors $\tilde{\beta}_i=(\tilde{\beta}_{i,1},\cdots,\tilde{\beta}_{i,N})^T$ with the entries $\tilde{\beta}_{i,j}=|j\rangle_{B_{i}}|\psi_{sym}\rangle_{1,\cdots,\hat{i},\cdots,n}$, 
where $|\psi_{sym}\rangle_{1,\cdots,\hat{i},\cdots,n}=\prod_{j\neq i}(\sum_{k=1}^{N}\frac{1}{\sqrt{N}}|k\rangle_{B_j})$ represents the totally symmetric state constructed with $n-1$ particles of Bob in the absence of the $i$-th one. The system's transitional state before the unitary operation $U^{\dagger}_{2}$ is therefore given as
\begin{eqnarray}
|\Omega\rangle&=&\sqrt{p_1}\alpha^T_1\Sigma_d\tilde{\beta}_1+\cdots+\sqrt{p_n}\alpha^T_n\Sigma_d\tilde{\beta}_n,
\label{B}
\end{eqnarray}
where $\langle \alpha_i,\alpha_j\rangle=M\delta_{i,j}$. If the input copies are not entangled with the coefficient matrix $\Sigma_d$ of the product states being also rank one, $|\Omega\rangle$ will naturally reduce to the tensor product of the state vectors of $1+n$ particles of Alice and Bob.  Due to the indistinguishableness of Bob's particles $B_i$, we can write $|k\rangle_{B_i}=|k\rangle_{B}$ (or $ b^{\dagger}_k|0\rangle_{B_i}=b^{\dagger}_k|0\rangle_{B}$ in terms of the creation operators of all modes $k$) so that $\tilde{\beta}_i=\tilde{\beta}$, for $i=1,\cdots,n$, in Eq. (\ref{B}) to reduce the coefficient matrix of the state $|\Omega\rangle$ to
\begin{eqnarray}
\Lambda(|\Omega\rangle)=\left(\begin{array}{c} \sqrt{p_1}~\Delta \\
\vdots\\
\sqrt{p_n}~\Delta
\end{array}\right),
\label{A}
\end{eqnarray}
where $\Delta$ is an $M\times N^{n}$ matrix in the form of
\begin{eqnarray}
\frac{1}{N^{(n-1)/2}}\left(\begin{array}{ccccccc}\sigma_1& \cdots & \sigma_1& & & & \\
& & &\sigma_2&\cdots &  \sigma_2& \\
& & & & & & \ddots \\
\end{array}\right).~~
\end{eqnarray}
The reduced density matrix, $\rho_A=\Lambda\Lambda^{\dagger}$, evolves under $U_{2}^{\dagger}$ of Alice to that of the target state $|\Psi_2\rangle$ in one subspace:
\begin{eqnarray}
\rho_{A, out}=U_{2}^{\dagger}~\rho_A U_{2}=\left(\begin{array}{cc} \Delta\Delta^{\dagger} & 0\\
0 & 0\\
\end{array}\right)=\left(\begin{array}{cc} \Sigma_d\Sigma^{\dagger}_d & 0\\
0 & 0\\
\end{array}\right).
\end{eqnarray}
The output state of the total system is, therefore, a graph state (see, e.g., \cite{graph}) of Alice's particle being simultaneously entangled in the same way to $n$ identical ones of Bob:
\begin{eqnarray}
|\Psi_{out}\rangle&=&\frac{1}{\sqrt{n!}}(U_{A,B_n}\cdots (U_{A,B_1}|\gamma_1\rangle_A|\gamma_2\rangle_{B_1})\cdots|\gamma_2\rangle_{B_n} \nonumber\\
&+& permutations~ with~ B_1, \cdots, B_n),
\end{eqnarray}
where $U_{A,B_{i}}|\gamma_1,\gamma_2\rangle_{A,B_{i}}=|\Psi_2\rangle_{A,B_{i}}$ is the joint unitary operation to generate the entanglement between Alice's particle and Bob's $i$-th particle, and $|\gamma_k\rangle$ the proper states in Alice and Bob's spaces. The joint unitary maps are commutative, $[U_{A, B_{i}},U_{A, B_{j}}]=0$, for any pair of $i$ and $j$, and $U_{A, B_{i}}=U_{B_{i}, A}$. To extract one copy of the target state out of this graph state, Bob only needs to select one of the particles and discard the remaining ones since the particles on his side are not entangled. In the whole process, the total amount of the classical information to be sent from Alice to Bob is at most one bit, telling him that her work has been done.

The minimum copy number of $|\Psi_1\rangle$ required to produce a target state is determined by $p^{opt}$ of our basic entanglement transformation strategy as $n_{min}=[max_k~\sigma_k^2/\lambda_k^2]+1$, where $[x]$ is the greatest integer 
less than $x$. This number gives the following result of manipulating finite copies of entangled state in analogy to the asymptotic result (in the limit of infinite copies) in Ref. \cite{lo-popescu}: if $n$ pairs of the input $|\Psi_{1}\rangle$ are manipulated by such LOCC strategy, the optimum probability of obtaining $nK$ target state $|\Psi_{2}\rangle$, where $n$ is any positive integer, 
is $1$ or $0$ when $K<1/n_{min}$ or $K>1/n_{min}$, respectively.

An interesting feature of this type of multi-copy entangled state manipulation is that the system's state is determined by the symmetry between the particles, as well as by the local unitary operations on them. There is a big difference caused by breaking the symmetry when Bob's particles $B_i$ are no longer identical. Now the $\tilde{\beta}_i$ in Eq. (\ref{B}) cannot be reduced to the same $\tilde{\beta}$, so the submatrix in each row of Eq. (\ref{A}) will be replaced by $\sqrt{p_i}~\Delta V_i$, where $V_i$ is an $N^n\times N^n$ permutation matrix of the basis vector of the $n$ {\it distinguishable} particles of Bob. Due to the existence of these permutation matrices, the eigenvalues of the reduced density matrix $\rho_A$ deviate from those of $\Sigma_d\Sigma^{\dagger}_d$, and any further transformation toward the target state will be impossible. 

\section{Conclusion}\label{section5}

In conclusion, we have investigated the transformations between bipartite entangled pure states from an implementation point of view. The strategy we propose answers an open question in \cite{lo-popescu}: how to achieve a better efficiency than the Schmidt projection of Bennett, {\it et al.} \cite{bennett}, which generates ME states without classical communication. We also examine the necessity of classical communication in realizing the deterministic transformations of a single copy of 
a bipartite entangled state \cite{nielsen, j-s}. To produce an entangled pair of bosons with certainty, we can extend this strategy by a direct sum scheme involving the bilateral actions to manipulate a definite number of any other pair of such particles with the same Schmidt rank.

\begin{acknowledgments}
The authors acknowledge the partial support by a grant from PSC-CUNY.
\end{acknowledgments}

\bibliographystyle{unsrt}

\begin{thebibliography}{99}
\bibitem{matrix}If $\Lambda$ is of rank one, $|\Psi_1\rangle$ will be a separable state.
\bibitem{notes1} $\lambda_i$ is mapped to $\sigma_i$ such that $\sigma_i\geq \sigma_{i+1}$. 
If not, the smallest $\lambda^2_{min(M,N)}$ will be scaled down to $p~\sigma^2_i$, where $\sigma_i\geq\sigma_{min(M,N)}$. 
Then the common factor $p$ will be smaller.
\bibitem{particle}The effective particle numbers after an SST of Alice is given as $Tr(\hat{n}_{A}\rho_{A_i})=p_i\langle0|\vec{a}\Sigma_d\Sigma^{\dagger}_d
\vec{a}^{\dagger}|0\rangle=p_i$, where $\hat{n}_{A}$ is the number operator and $\vec{a}^{\dagger}=(a^{\dagger}_{1},\cdots, a^{\dagger}_{M})^T$.
\end{thebibliography}

\end{document}